\documentclass[twocolumn,preprintnumbers,amsmath,amssymb]{revtex4}

\usepackage{graphicx}
\usepackage{dcolumn}
\usepackage{bm}
\usepackage{theorem}
\usepackage{color}
\usepackage{multirow}
\usepackage{amssymb}

{\theorembodyfont{\normalfont\it}
\theoremheaderfont{\normalfont\bf}
\newtheorem{thm}{ Theorem}
\newtheorem{dfn}[thm]{ Definition}
\newtheorem{lmm}[thm]{ Lemma}

\newtheorem{crl}[thm]{ Corollary}
\newtheorem{asm}[thm]{ Assumption}
\newtheorem{prp}[thm]{ Proposition}
\newtheorem{cjt}[thm]{ Conjecture}}

{\theorembodyfont{\normalfont}
\theoremheaderfont{\normalfont\it}
\newtheorem{prf}{ Proof:}}

{\theorembodyfont{\normalfont}
\theoremheaderfont{\normalfont\it}
}

{\theorembodyfont{\normalfont}
\theoremheaderfont{\normalfont\it}
}

{\theorembodyfont{\normalfont}
\theoremheaderfont{\normalfont\it}
}

{\theorembodyfont{\normalfont}
\theoremheaderfont{\normalfont\it}
}

{\theorembodyfont{\normalfont}
\theoremheaderfont{\normalfont\it}
}

{\theorembodyfont{\normalfont}
\theoremheaderfont{\normalfont\it}
\newtheorem{rmk}{ Remark.}}

\newcommand{\bra}[1]{\mbox{$\langle#1|$}}

\newcommand{\ket}[1]{\mbox{$|#1\rangle$}}

\newcommand{\proj}[1]{\mbox{$\ket{#1}\!\bra{#1}$}}

\newcommand{\alg}[1]{\begin{align}#1\end{align}}

\newcommand{\ca}[1]{{\mathcal #1}}
\newcommand{\mbb}[1]{{\mathbb #1}}

\newcommand{\bthm}[1]{\begin{thm}\label{thm:#1}}
\newcommand{\ethm}{\end{thm}}

\newcommand{\rThm}[1]{Theorem \ref{thm:#1}}
\newcommand{\blmm}[1]{\begin{lmm}\label{lmm:#1}}
\newcommand{\elmm}{\end{lmm}}

\newcommand{\bdfn}[1]{\begin{dfn}\label{dfn:#1}}
\newcommand{\edfn}{\end{dfn}}

\newcommand{\basm}[1]{\begin{asm}\label{asm:#1}}
\newcommand{\easm}{\end{asm}}
\newcommand{\bprp}[1]{\begin{prp}\label{prp:#1}}
\newcommand{\eprp}{\end{prp}}

\newcommand{\rPrp}[1]{Proposition \ref{prp:#1}}
\newcommand{\bcrl}[1]{\begin{crl}\label{crl:#1}}
\newcommand{\ecrl}{\end{crl}}

\newcommand{\rCrl}[1]{Corollary \ref{crl:#1}}
\newcommand{\bcjt}[1]{\begin{cjt}\label{cjt:#1}}
\newcommand{\ecjt}{\end{cjt}}

\newcommand{\bprf}{\begin{prf}}
\newcommand{\eprf}{\end{prf}}
\newcommand{\brmk}{\begin{rmk}}
\newcommand{\ermk}{\end{rmk}}
\newcommand{\laeq}[1]{\label{eq:#1}}
\newcommand{\req}[1]{(\ref{eq:#1})}

\newcommand{\QED}{\hfill$\blacksquare$}
\newcommand{\lsec}[1]{\label{sec:#1}}

\newcommand{\rSec}[1]{Section \ref{sec:#1}}
\newcommand{\lapp}[1]{\label{app:#1}}

\newcommand{\rApp}[1]{Appendix \ref{app:#1}}

\newcommand{\sR}{s_{\scalebox{0.45}{$R$}}}
\newcommand{\sL}{s_{\scalebox{0.45}{$L$}}}
\newcommand{\sz}{s_{\scalebox{0.45}{$0$}}}

\newcommand{\bsR}{{\bar s}_{\scalebox{0.45}{$R$}}}
\newcommand{\bsL}{{\bar s}_{\scalebox{0.45}{$L$}}}
\newcommand{\bsz}{{\bar s}_{\scalebox{0.45}{$0$}}}

\newcommand{\bitem}{\begin{itemize}}
\newcommand{\entem}{\end{itemize}}
\newcommand{\benum}{\begin{enumerate}}
\newcommand{\ennum}{\end{enumerate}}

\newcommand{\otm}{\otimes}

\begin{document}


\title{Superdense Coding in Resource Theory of Asymmetry}

\author{Eyuri Wakakuwa}
\email{e.wakakuwa@gmail.com}
\affiliation{Department of Communication Engineering and Informatics, Graduate School of Informatics and Engineering, The University of Electro-Communications, Tokyo 182-8585, Japan}


\begin{abstract}
We analyze the task of encoding classical information into a quantum system under the restriction by symmetry.
Motivated by an analogy between the resource theories of asymmetry and entanglement,
we ask whether an analog of superdense coding is possible in the former.
I.e., we investigate whether the classical information capacity of an asymmetric state can be strictly larger than that of any symmetric state whereas the latter is a strictly positive constant.
We prove that this is possible if and only if the unitary representation of the symmetry is non-Abelian and reducible.
The result provides an information-theoretical classification of symmetries of quantum systems.
We also discuss the possibility of superdense coding in other resource theories.
\end{abstract}

                       
                                                      
\maketitle

\section{Introduction}
Asymmetry of quantum states plays the role of resources for information processing tasks when the operations on the system are restricted by symmetry. 
Examples of the tasks range from quantum communication \cite{bartlett09} and quantum metrology \cite{hall12} to reference frame sharing \cite{bartlett07} and thermodynamic work extraction \cite{vaccaro2008tradeoff}.
Fundamental limitations on the transformation of quantum states under the symmetry restriction have been investigated in Refs.~\cite{gour2008resource,toloui2012simulating,marvian2013theory,marvian2014asymmetry,marvian14,marvian2014modes}, and quantification of asymmetry of quantum states has been addressed in \cite{gilad08,toloui2011constructing,piani16,takagi2019skew}.
These researches are referred to as the {\it resource theory of asymmetry}. 
Theoretically, the resource theory of asymmetry is a particular case of a general formalism called the {\it quantum resource theory} \cite{chitambar2019quantum}. 
This formalism also applies to other physical properties such as coherence \cite{streltsov2017colloquium}, athermality \cite{brandao2013resource}, purity \cite{horodecki2003reversible}, non-Gaussianity (see e.g.~\cite{braunstein2005quantum,lami2018gaussian}), and most notably, entanglement \cite{plenio2014introduction, horodecki2009quantum}.
Once we have an information processing task for which one physical property (e.g., entanglement) is useful as a resource, we could find, by formal analogy, a task for which another property (e.g., asymmetry) plays the role of the resource.

\begin{figure}[t]
\begin{center}
\includegraphics[bb={0 50 513 468}, scale=0.3]{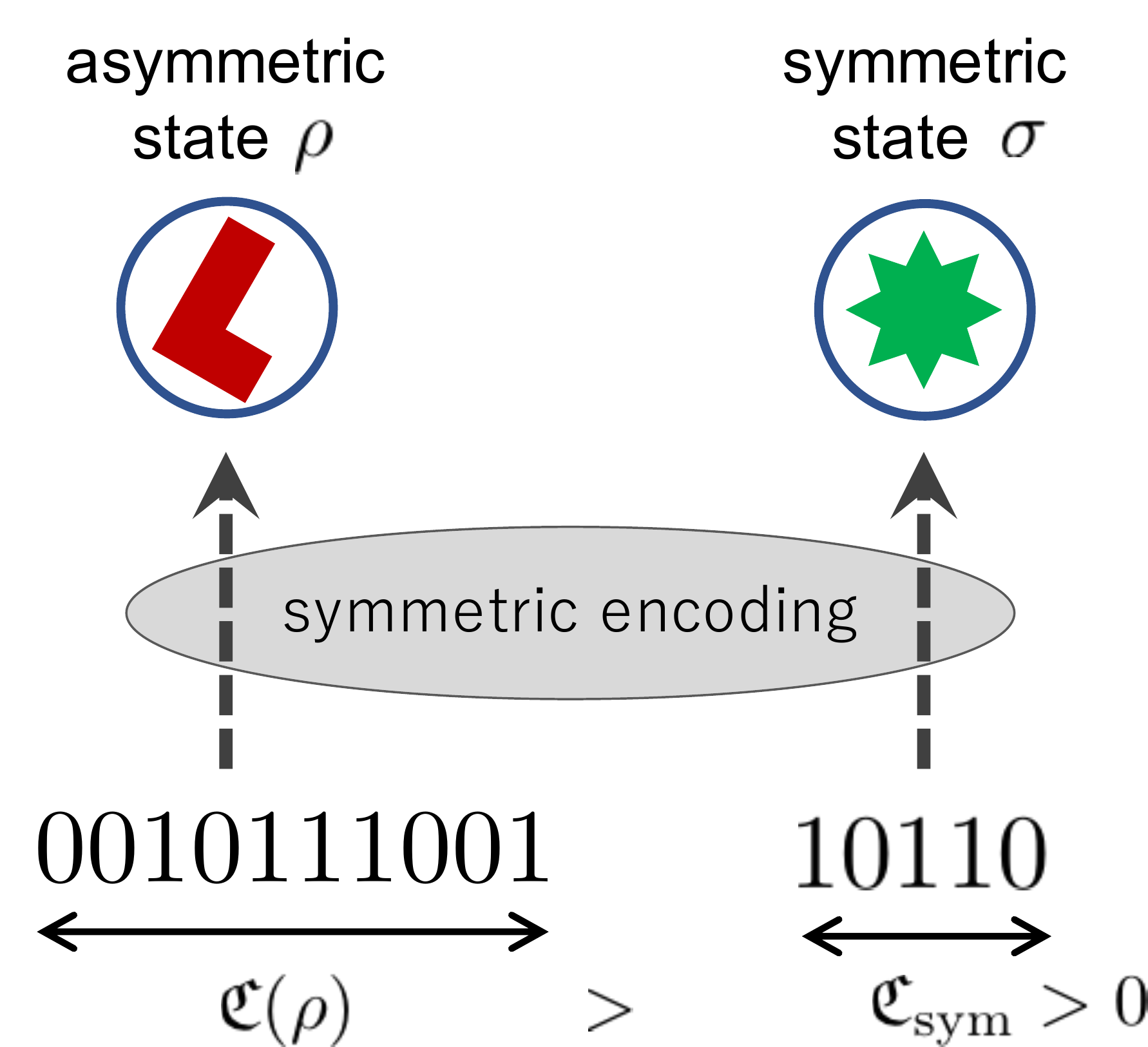}
\end{center}
\caption{
Superdense coding in the resource theory of asymmetry is depicted.
The bit arrays are classical information that are to be encoded into quantum systems.
The quantum systems are initially in an asymmetric state $\rho$ or in a symmetric state $\sigma$. 
The function $\mathfrak{C}$ denotes the symmetry-restricted classical information capacity of the state.
$\mathfrak{C}_{\rm sym}$ is a constant that does not depend on $\sigma$.
}
\label{fig:A}
\end{figure}

In this paper, we adopt the task of superdense coding in the resource theory of entanglement and investigate the possibility of its analog in the resource theory of asymmetry.
We consider a scenario in which one encodes classical information into a quantum system by an operation restricted by symmetry.
The maximum amount of classical information that can be encoded under this restriction depends on the initial state of the system.
In analogy with superdense coding in the resource theory of entanglement, we define that superdense coding is possible if the classical information capacity of an asymmetric state is strictly larger than that of any symmetric state whereas the latter is a strictly positive constant (see Figure \ref{fig:A}).
We prove that this relation holds if and only if the unitary representation of the symmetry of the system is non-Abelian and reducible.
Consequently, superdense coding is possible in systems, e.g., with ${\rm SU}(2)$ symmetry, whereas it is not possible in the case, e.g., of $ {\rm U}(1)$ symmetry.
The result thus provides an information-theoretical classification of symmetries of quantum systems.
For the simplicity of analysis, we consider an asymptotic limit of infinitely many copies and vanishingly small error.

This paper is organized as follows.
In \rSec{def}, we describe the framework of the resource theory of asymmetry.
In \rSec{formres}, we formulate the problem and present the main results.
In \rSec{ORT}, we discuss the possibility of superdense coding in other resource theories.
Conclusions are given in \rSec{conc}.
Throughout this paper, $\log{t}$ represents the base $2$ logarithm of $t$.

\section{Resource Theory of Asymmetry}
\label{sec:def}

We describe the framework of the resource theory of asymmetry.

\subsection{Symmetry of Quantum States}

Consider a quantum system $S$ described by a Hilbert space ${\ca H}^S$ with dimension $d_S\:(<\infty)$.
The set of normalized density operators on ${\ca H}^S$ is denoted by ${\cal S}({\ca H}^S)$.
Consider a symmetry group $G$ with the  unitary representation $\mathfrak{U}_G\equiv\{U_g\}_{g\in G}$ on ${\ca H}^S$. A state $\sigma\in{\cal S}({\ca H}^S)$ is said to be {\it symmetric} if it satisfies
$
U_g\sigma U_g^\dagger=\sigma
$
for any $g\in G$,
and {\it asymmetric} otherwise. We denote the set of symmetric states by ${\ca S}_{\rm sym}(S,\mathfrak{U}_G)$.
This definition can naturally be generalized to a system composed of $n$ duplicates of a system $S$, which is denoted by $S^n$.
We introduce notations ${\vec g}:=(g_1,\cdots,g_n)$ and $U_{\vec g}:=U_{g_1}\otimes\cdots\otimes U_{g_n}$ for $g_i\in G\;(i=1,\cdots,n)$. 
Symmetric states are those that satisfy the condition $U_{\vec g}\sigma U_{\vec g}^\dagger=\sigma$ for any ${\vec g}\in G^{\times n}$ (see Appendix C in \cite{wakakuwa2017symmetrizing}).

A quantum system with symmetry is represented by a Hilbert space that is decomposed into a direct-sum-product form \cite{bartlett07}. 
The decomposition is also represented by an embedded form (see Appendix B.1 in Ref.~\cite{wakakuwa2021one} for the detail), which would be useful to simplify the notations.
Consider a symmetry group $G$ with the unitary representation $\mathfrak{U}_G\equiv\{U_g\}_{g\in G}$ on ${\ca H}^S$. There exist three Hilbert spaces ${\ca H}^{\sz}$, ${\ca H}^{\sL}$, ${\ca H}^{\sR}$ and a linear isometry $\Gamma:{\ca H}^S\rightarrow{\ca H}^{\sz}\otimes{\ca H}^{\sL}\otimes{\ca H}^{\sR}$ that satisfy the following conditions: 
First, $\Gamma\Gamma^\dagger=\sum_{q\in Q}\proj{q}^{\sz}\otimes I_q^{\sL}\otimes I_q^{\sR}$,
where $\{|q\rangle\}_{q\in Q}$ is a fixed orthonormal basis of ${\ca H}^{\sz}$, and $I_q^{\sL}$, $I_q^{\sR}$ are projectors on subspaces ${\ca H}_q^{\sL}\subseteq{\ca H}^{\sL}$ and ${\ca H}_q^{\sR}\subseteq{\ca H}^{\sR}$, respectively, for each $q$. Second, for any $g\in G$, the unitary $U_g$ is decomposed by $\Gamma$ as 
$
\Gamma U_g\Gamma^\dagger=\sum_{q\in Q}\proj{q}^{\sz}\otimes u_{g,q}^{\sL}\otimes I_q^{\sR}
$,
where $\{u_{g,q}\}_{g\in G}$ is an irreducible unitary representation of $G$ on ${\ca H}_q^{\sL}\subseteq{\ca H}^{\sL}$ for each $q$. The linear isometry $\Gamma$ satisfying the above conditions is uniquely determined from $\mathfrak{U}_G$ up to changes of the basis. 
Schur's lemma implies that a symmetric state $\sigma\in{\ca S}_{\rm sym}(S,\mathfrak{U}_G)$ is decomposed by $\Gamma$ as
$
\Gamma\sigma\Gamma^\dagger=\sum_{q\in Q}r_q\proj{q}^{\sz}\otimes\pi_q^{\sL}\otimes \sigma_q^{\sR}
$,
where $\{r_q\}_{q\in Q}$ is a probability distribution on $Q$, $\pi_q$ is the maximally mixed state on ${\ca H}_q^{\sL}$ and $\sigma_q$ is a state on ${\ca H}_q^{\sR}$ for each $q\in Q$. 
For the system composed of $n$ duplicates of $S$, it is convenient to introduce notations $\vec{q}:=(q_1,\cdots, q_n)\in Q^{\times n}$,
$
\ca{H}_{\vec{q}}^{\bsL}:=\ca{H}_{q_1}^{\sL}\otm\cdots\otm\ca{H}_{q_n}^{\sL}
$,
$
\ca{H}_{\vec{q}}^{\bsR}:=\ca{H}_{q_1}^{\sR}\otm\cdots\otm\ca{H}_{q_n}^{\sR}
$
and 
$\ca{H}^{\bsz}:=(\ca{H}^{\sz})^{\otm n}$.
A state $\sigma_n$ on $S^n$ is symmetric if and only if it is decomposed by $\Gamma^{\otimes n}$ as 
$
(\Gamma^{\otimes n})\sigma(\Gamma^{\dagger \otimes n})=\sum_{\vec{q}\in Q^{\times n}}r_{\vec{q}}\proj{\vec{q}}^{\bsz}\otimes\pi_{\vec{q}}^{\bsL}\otimes \sigma_{\vec{q}}^{\bsR}
$, 
with $\{r_{\vec{q}}\}_{\vec{q}\in Q^{\times n}}$ being a probability distribution, $\pi_{\vec{q}}^{\bsL}$ being the maximally mixed state on $\ca{H}_{\vec{q}}^{\bsL}$ and $\sigma_{\vec{q}}\in\ca{S}({\ca H}_{\vec{q}}^{\bsR})$.
In the rest of this paper, we omit $\Gamma$ and represent all states and operators on $S$ in the embedded form.
 
Abbreviating ${\rm dim}\ca{H}_q^{\sL}$ and ${\rm dim}\ca{H}_q^{\sR}$ as $d_q^{\sL}$ and $d_q^{\sR}$, respectively,
the total dimension of the system is calculated to be
$
d_S=\sum_{q\in Q}d_q^{\sL}d_q^{\sR}
$.
The unitary representation $\mathfrak{U}_G$ of $G$ on $\ca{H}^S$ is Abelian if and only if $d_q^{\sL}=1$ for all $q\in Q$, or equivalently, iff it holds that
\alg{
 d_S=\sum_{q\in Q}d_q^{\sR}.
 \laeq{dimAbelian}
 }
The representation $\mathfrak{U}_G$ is irreducible if and only if $|Q|=1$ and $d_q^{\sR}=1$,
which is equivalent to the condition that
 \alg{
\sum_{q\in Q}d_q^{\sR}=1.
 \laeq{dimirreducible}
 }

\subsection{Symmetry of Quantum Operations}

Any quantum operation on system $S$ is represented by a completely positive trace-preserving (CPTP) map $\ca{K}$ on ${\mathcal S}({\mathcal H}^S)$, which is described by a (non-unique) set of linear operators $\{K_l\}_l$ on $\ca{H}^S$ such that
$\ca{K}(\cdot)=\sum_lK_l(\cdot) K_l^\dagger$ and $\sum_lK_l^\dagger K_l= I$ (see e.g.~\cite{nielsentext}).
We say that a CPTP map $\ca{K}$ on ${\mathcal S}({\mathcal H}^S)$ is {\it asymmetry-nongenerating} (AN) if $\ca{K}(\sigma)\in{\ca S}_{\rm sym}(S,\mathfrak{U}_G)$ for any $\sigma\in {\ca S}_{\rm sym}(S,\mathfrak{U}_G)$, and {\it symmetry-preserving} (SP) if $\ca{K}^\dagger(\sigma')/{\rm Tr}[\ca{K}^\dagger(\sigma')]\in{\ca S}_{\rm sym}(S,\mathfrak{U}_G)$ for any $\sigma'\in {\ca S}_{\rm sym}(S,\mathfrak{U}_G)$ as well, where $\ca{K}^\dagger$ is the adjoint map of $\ca{K}$.
In addition, we say that an operation $\ca{K}$ is {\it strongly asymmetry-nongenerating} (SAN) if there exists at least one representation $\{K_l\}_l$  so that
\alg{
\forall l, \forall \sigma\in {\ca S}_{\rm sym}(S,\mathfrak{U}_G);
\;
\frac{K_l\sigma K_l^\dagger}{{\rm Tr}[K_l\sigma K_l^\dagger]}\in{\ca S}_{\rm sym}(S,\mathfrak{U}_G),
\laeq{symnongen}
}
and {\it strongly symmetry-preserving} (SSP) if, in addition to \req{symnongen}, it holds that 
\alg{
\forall l, \forall \sigma'\in {\ca S}_{\rm sym}(S,\mathfrak{U}_G);
\;
\frac{K_l^\dagger\sigma' K_l}{{\rm Tr}[K_l^\dagger\sigma' K_l]}\in{\ca S}_{\rm sym}(S,\mathfrak{U}_G).
\laeq{symnongen2}
}

Suppose that our ability to perform operations on system $S$ is restricted by symmetry.
It would be natural to assume that asymmetric states cannot be generated from symmetric states.
Thus, we consider one of the above four classes as the set of operations allowed under the restriction by symmetry, which we denote by ${\ca O}_{\rm sym}(S,\mathfrak{U}_G)$. 
A {\it symmetric operation} refers to an operation that is either (i) preparation of the system in a symmetric state or (ii) an operation that belongs to ${\ca O}_{\rm sym}(S,\mathfrak{U}_G)$.
It is straightforward that the set of symmetric states is closed under symmetric operations and that the set of symmetric operations is closed under sequential compositions. 
Due to the convexity of ${\ca S}_{\rm sym}(S,\mathfrak{U}_G)$, any SSP operation is also AN, SP, and SAN.
For the unitary operations, AN coincide with SAN and SP coincide with SSP.
The results presented in this paper do not depend on which of the four classes we choose.

Instead of the above four classes, one may also consider covariant operations \cite{bartlett2003entanglement,gour2008resource,toloui2012simulating,marvian2013theory,marvian2014asymmetry,marvian14,marvian2014modes} or operations that keep the symmetric twirling operation invariant \cite{korzekwa2019encoding} as symmetric operations.
In this paper, we will not discuss which one of them is the ``natural'' choice, because
it should be decided depending on each physical context.
We remark that the classes AN, SAN, and SSP are natural generalizations of MIO (maximally incoherent operations), IO (incoherent operations), and SIO (strictly incoherent operations) in the resource theory of coherence \cite{chitambar2016critical}.

\section{Superdense Coding}
\lsec{formres}

We present formulation of superdense coding in the resource theory of asymmetry and describe the main result.
A proof of the result will be provided in the following subsections.

\subsection{Formulation and Main Results}

Let us first recall superdense coding in the resource theory of entanglement \cite{bennett1992communication,horodecki2001classical}.
Consider encoding of a classical message into a composite quantum system $AB$ with a fixed dimension $d_Ad_B$. 
If any operation on $AB$ is allowed for encoding, at most $\log{d_Ad_B}$ bits of classical information can be encoded.
Suppose, however, that the encoding operation is somehow restricted to operations on system $A$ alone. 
In this case, the maximum amount of classical information that can be encoded into system $AB$ depends on its initial state.
If the initial state is separable, the maximum encodable information is equal to $\log{d_A}$.
On the other hand, when the initial state is sufficiently strongly entangled, it can be strictly larger than $\log{d_A}$.

To formulate an analog of superdense coding in the resource theory of asymmetry, 
consider a scenario in which the sender aims at transmitting $nR$ bits of classical information $X$ by encoding it into a quantum system $S^n=S_1\cdots S_n$ and sending the system to the receiver.
Each system $S_i$ is initially in a state $\rho$.
We assume that the encoding operation $\ca{E}$ is restricted to be symmetric ones, in which case it is described by a set of symmetric CPTP maps $\{\ca{E}_x\}_{x=1,\cdots,2^{nR}}$.
A decoding operation is described by a measurement $\ca{M}$ on $S^n$, represented in terms of a set of positive semidefinite operators $\{M_{x}\}_{x=1,\cdots,2^{nR}}$ such that $\sum_{x=1,\cdots,2^{nR}}M_x\leq I$.
The probability of correctly decoding $X$ when the value of the message is $x$ is given by
$
{\rm Pr}(\text{correct}|X=x)={\rm Tr}[M_x\ca{E}_x(\rho^{\otm n})].
$
Thus, the maximum probability of error is defined by
$
\varepsilon(\ca{E},\ca{M},\rho^{\otm n})
:=
\max_{x\in\{1,\cdots,2^{nR}\}}[1-{\rm Tr}[M_x\ca{E}_x(\rho^{\otm n})]].
$
The {\it symmetry-restricted classical information capacity} of a state $\rho$ is defined as the supremum of the rate $R$ such that the maximum probability of error can be made arbitrarily small for any sufficiently large $n$ by properly choosing the encoding operation and the decoding measurement. 
We denote this capacity by $\mathfrak{C}(\rho)$.
Since the set of symmetric operations is closed under sequential compositions, 
the function $\mathfrak{C}$ is monotonically nonincreasing under symmetric operations.

Our interest is on the maximal values of the symmetry-restricted classical information capacities of asymmetric states and symmetric states.
Analogously to superdense coding in the resource theory of entanglement, we focus on whether or not the former is strictly larger than the latter whereas the latter is a strictly positive constant.

\bdfn{superdense}
{\it Superdense coding is possible} on system $S$ with symmetry $G$  if there exists a constant $\mathfrak{C}_{\rm sym}>0$ and it holds that
\alg{
\max_{\rho\in\ca{S}(\ca{H}^S)}\mathfrak{C}(\rho)
>
\mathfrak{C}_{\rm sym}
=
\mathfrak{C}(\sigma)
}
for any $\sigma\in\ca{S}_{\rm sym}(S,\mathfrak{U}_{G})$.
\edfn

\noindent
The main result of this paper is the following theorem, which is applicable to any of AN, SP, SAN and SSP:
\bthm{densecoding}
Superdense coding is possible on system $S$ with symmetry $G$ if and only if the unitary representation $\mathfrak{U}_G$  of $G$ on $\ca{H}^S$ is non-Abelian and reducible.
\ethm

\noindent
A proof of \rThm{densecoding} is provided in the following subsections. For the properties of quantum entropies that are used in the proof, see e.g. \cite{nielsentext,hayashitext,cover05,wildetext}.

An operation $\ca{K}$ on system $S$ is said to be {\it covariant} if it commutes with the action of $U_g$, i.e., if it satisfies $\ca{K}(U_g(\cdot)U_g^\dagger)=U_g\ca{K}(\cdot)U_g^\dagger$, for any $g\in G$ \cite{bartlett2003entanglement,gour2008resource,toloui2012simulating,marvian2013theory,marvian2014asymmetry,marvian14,marvian2014modes}.
One can also consider encoding of classical information into a quantum system by covariant operations.
Superdense coding is possible under similar conditions as in \rThm{densecoding} (see \rApp{capCE}).
Ref.~\cite{korzekwa2019encoding} considered a similar task in which one encodes classical information into a quantum system by operations that keep the symmetric twirling operation invariant.

\subsection{Capacity of Asymmetric States}
The first step toward the proof of \rThm{densecoding} is to obtain a lower bound on the symmetry-restricted classical information capacity of arbitrary states:

\bprp{MICofAsym}
For any symmetry $G$ with the unitary representation $\mathfrak{U}_G$ on $\ca{H}^S$ and any state $\rho\in\ca{S}(\ca{H}^S)$, it holds that
\alg{
\mathfrak{C}(\rho)
\geq
H(\{p_q\})+\sum_{q\in Q}p_q\log{d_q^{\sL}d_q^{\sR}}-H(\rho),
\laeq{HH}
}
where $p_q:={\rm Tr}[\bra{q}^{\sz}\rho\ket{q}^{\sz}]$.
The symbol $H$ denotes both the Shannon entropy and the von Neumann entropy, defined by $H(\{p_q\}):=-\sum_qp_q\log{p_q}$ and $H(\rho):=-{\rm Tr}[\rho\log{\rho}]$.
\eprp

\bprf
Fix arbitrary $\epsilon,\delta>0$ and choose sufficiently large $n$.
Let $\ca{T}_{n,\delta}\in\ca{Q}^{\times n}$ be the $\delta$-strongly typical set with respect to $\{p_q\}_q$,
and define a projector
$
\Pi_{n,\delta}^0:=
\sum_{\vec{q}\in\ca{T}_{n,\delta}}
\proj{\vec{q}}^{\bsz}\otm 
I_{\vec{q}}^{\bsL}\otm I_{\vec{q}}^{\bsR}$.
Let $W$ be a unitary on $S^n$ that is decomposed into
$
W=\sum_{\vec{q}\in Q^{\times n}}\proj{\vec{q}}^{\bsz}\otm u_{\vec{q}}^{\bsL}\otm v_{\vec{q}}^{\bsR}
$,
where $u_{\vec{q}}$ and $v_{\vec{q}}$ are unitaries on $\ca{H}_{\vec{q}}^{\bsL}$ and $\ca{H}_{\vec{q}}^{\bsR}$, respectively.
It is straightforward to verify that $W$ is symmetry-preserving.
Using this unitary, we define
$
\rho_{W,n}:=W\rho^{\otm n}W^\dagger
$.
Due to the property of the typical set, it holds that
$
{\rm Tr}[\Pi_{n,\delta}^0\rho_{W,n}]
\geq1-\epsilon.
$
Suppose that $u_{\vec{q}}$ and $v_{\vec{q}}$ in the definition of $W$ are chosen randomly according to the unitary invariant (Haar) measure, independently for each $\vec{q}$.
The averaged state over the unitaries is given by
$
\bar{\rho}_n:=\mbb{E}_{W}[\rho_{W,n}]
=
\sum_{\vec{q}\in Q^{\times n}}p_{q_1}\cdots p_{q_n}\proj{\vec{q}}^{\bsz}\otm \pi_{\vec{q}}^{\bsL}\otm \pi_{\vec{q}}^{\bsR}
$.
Denoting the R.H.S. of \req{HH} by $\hat{D}(\rho)$,
it follows that
$
\Pi_{n,\delta}^0\bar{\rho}_n\Pi_{n,\delta}^0
\leq
2^{-n[\hat{D}(\rho)-\delta]}
\cdot \Pi_{n,\delta}^0
$.
We also define a projector $\Pi_{W,n,\delta}:=W\Pi_{n,\delta}W^\dagger$ for each $W$, where $\Pi_{n,\delta}$ is the projection onto the $\delta$-typical subspace of $(\ca{H}^S)^{\otm n}$ with respect to $\rho$.
It should be noted that ${\rm Tr}[\Pi_{W,n,\delta}]\leq2^{n[H(\rho)+\delta]}$ and ${\rm Tr}[\Pi_{W,n,\delta}\rho_{W,n}]\geq1-\epsilon$.

To prove the existence of an encoding operation and a decoding measurement with a small error, we fix $R\equiv \hat{D}(\rho)-3\delta$ and apply the packing lemma (Corollary 15.5.1 in \cite{wildetext}). 
It follows that there exist a set of symmetry-preserving unitaries $\{W_x\}_{x=1}^{2^{nR}}$ and a POVM $\{\Lambda_x\}_{x=1}^{2^{nR}}$ such that
$
2^{-nR}\sum_{x=1}^{2^{nR}}{\rm Tr}[\Lambda_x\rho_{W_x,n}]
\geq1-4(\epsilon+2\sqrt{\epsilon})-8\cdot2^{-n\delta}.
$
The R.H.S. of this inequality is greater than $1-13\sqrt{\epsilon}$ for sufficiently large $n$.
We construct the protocol with rate $R$ and error $13\sqrt{\epsilon}$ by the encoding operations $\ca{E}_x(\cdot)=W_x(\cdot)W_x^\dagger$ and the decoding measurement $M_x=\Lambda_x\:(x=1,\cdots,2^{nR})$.
Since $\epsilon,\delta>0$ can be arbitrarily small, this completes the proof of Inequality \req{HH}.
\QED
\eprf

\noindent
The lower bound in \rPrp{MICofAsym} coincides with the dimension of the system for an asymmetric state in a particular form:

\bcrl{MICofAsym}
For any symmetry $G$ with the unitary representation $\mathfrak{U}_G$ on $\ca{H}^S$, it holds that
\alg{
\max_{\rho\in\ca{S}(\ca{H}^S)}\mathfrak{C}(\rho)
=
\log{d_S}.
}
\ecrl

\bprf
The converse part $\sup_{\rho}\mathfrak{C}(\rho)\leq\log{d_S}$ immediately follows from the Holevo bound \cite{holevo73}.
The direct part $\sup_{\rho}\mathfrak{C}(\rho)\geq\log{d_S}$ follows from \rPrp{MICofAsym}.
Note that the R.H.S. of Inequality \req{HH} is equal to $\log{d_S}$ for any pure state $\ket{\psi}$ in the form of
$\ket{\psi}=\sum_q\sqrt{d_q^{\sL}d_q^{\sR}/d_S}\ket{q}^{\sz}\ket{\psi_q}^{\sL\sR}$,
where $\ket{\psi_q}$ is a normalized state vector in $\ca{H}_q^{\sL}\otm\ca{H}_q^{\sR}$ for each $q$.
\QED
\eprf

\subsection{Capacity of Symmetric States}
The second step for the proof of \rThm{densecoding} is to derive the symmetry-restricted classical information capacity of symmetric states.
This completes the proof of \rThm{densecoding} when combined with the Abelian condition \req{dimAbelian}, the irreducibility condition \req{dimirreducible} and \rCrl{MICofAsym}.

\bprp{MICofSym}
For any symmetry $G$ with the unitary representation $\mathfrak{U}_G$ on $\ca{H}^S$ and any symmetric state $\sigma\in\ca{S}_{\rm sym}(S,\mathfrak{U}_G)$, it holds that
\alg{
\mathfrak{C}(\sigma)
=
\log{\left(\sum_{q\in Q}d_q^{\sR}\right)}.
\laeq{MICofSym}
}
\eprp

\bprf

To prove the direct part $\mathfrak{C}(\sigma)\geq\log{(\sum_qd_q^{\sR})}$, we show that $\log{(\sum_qd_q^{\sR})}$ bits of classical information can be encoded by a symmetric operation on system $S$.
 Let $\{\ket{e_{r|q}}\}_{r=1}^{d_q^{\sR}}$ be an orthonormal basis of $\ca{H}_q^{\sR}$ for each $q\in Q$,
and consider symmetric states $\varsigma_{q,r}=\proj{q}^{\sz}\otm\pi_{q}^{\sL}\otm\proj{e_{r|q}}^{\sR}\:(1\leq r\leq d_q^{\sR},q\in Q)$.
The supports of these states are orthogonal for $(q,r)\neq(q',r')$.
Thus, $\log{(\sum_qd_q^{\sR})}$ bits of classical information can be encoded into the system $S$ by preparing it in the state $\varsigma_{q,r}$, depending on the message.

Next, we prove the converse part $\mathfrak{C}(\sigma)\leq\log{(\sum_qd_q^{\sR})}$.
Suppose that $R<\mathfrak{C}(\sigma)$. 
By definition, for any $\epsilon>0$ and sufficiently large $n$, there exist a set of symmetric encoding operations $\ca{E}\equiv\{\ca{E}_x\}_{x=1,\cdots,2^{nR}}$ and a decoding measurement $\ca{M}\equiv\{M_x\}_{x=1,\cdots,2^{nR}}$ that satisfy the small error condition
$
\varepsilon(\ca{E},\ca{M},\rho^{\otm n})\leq\epsilon
$.
The state after the encoding operation is represented by the density operator
\alg{
\varsigma_n:=2^{-nR}\sum_{x=1}^{2^{nR}}\proj{x}^X\otm\ca{E}_x(\sigma^{\otm n}).
\laeq{eq15}
}
The state after the decoding measurement is represented by $\hat{\ca{M}}(\varsigma_n)$,
where $\hat{\ca{M}}:S^n\rightarrow\hat{X}$ is a map defined by $\hat{\ca{M}}(\tau)={\rm Tr}[M_x(\tau)]\cdot\proj{x}^{\hat{X}}$.
We introduce a state $\Theta_n:=2^{-nR}\sum_{x=1,\cdots,2^{nR}}\proj{x}^X\otm\proj{x}^{\hat{X}}$.
The small error condition implies that $\|\hat{\ca{M}}_n(\varsigma_n)-\Theta_n\|_1\leq\epsilon$.
Due to Fano's inequality (see e.g.~\cite{cover05}), the mutual information between $X$ and $\hat{X}$ is evaluated as
$
nR
\!=\!
I(X\! :\! \hat{X})_{\Theta_n}
\!\leq\!
I(X\! :\! \hat{X})_{\hat{\ca{M}}(\varsigma_n)}+nR\eta(\epsilon)
$,
where $\eta$ is a function that satisfies $\lim_{\epsilon\rightarrow0}\eta(\epsilon)=0$.
The monotonicity of quantum mutual information implies
$
I(X\! :\! \hat{X})_{\hat{\ca{M}}(\varsigma_n)}
\leq
I(X\! :\! S^n)_{\varsigma_n}
=
H(S^n)_{\varsigma_n}- H(S^n|X)_{\varsigma_n}
$.
As we prove below, the entropies of the state $\varsigma_n$ are calculated to be
$
H(S^n|X)_{\varsigma_n}
\geq
\sum_{\vec{q}}\mu_{\vec{q}}\log{d_{\vec{q}}^{\bsL}}
$
and
$
H(S^n)_{\varsigma_n}
\leq
H(\{\mu_{\vec{q}}\}_{\vec{q}})+\sum_{\vec{q}}\mu_{\vec{q}}
\log{d_{\vec{q}}^{\bsL}d_{\vec{q}}^{\bsR}}
$,
where $\{\mu_{\vec{q}}\}_{\vec{q}\in Q^{\times n}}$ is a probability distribution.
Thus, we arrive at
$
nR
\leq
\sum_{\vec{q}}\mu_{\vec{q}}\log{(d_{\vec{q}}^{\bsR}/\mu_{\vec{q}})}+nR\eta(\epsilon)
$.
Defining a probability distribution $\{\nu_{\vec{q}}\}_{\vec{q}}$ by
$
\nu_{\vec{q}}=d_{\vec{q}}^{\sR}/\sum_{\vec{q}}d_{\vec{q}}^{\bsR}
$,
and noting that $\sum_{\vec{q}}d_{\vec{q}}^{\bsR}=(\sum_{q}d_q^{\sR})^n$,
we have
$
\sum_{\vec{q}}\mu_{\vec{q}}\log{(d_{\vec{q}}^{\bsR}/\mu_{\vec{q}})}
=
-D(\{\mu_{\vec{q}}\}\|\{\nu_{\vec{q}}\})+n\log{(\sum_qd_q^{\sR})}
\leq
n\log{(\sum_qd_q^{\sR})}.
$
Here, $D$ is the Kullback-Leibler divergence defined by $D(\{\mu_{\vec{q}}\}\|\{\nu_{\vec{q}}\}):=\sum_{\vec{q}}\mu_{\vec{q}}\log{(\mu_{\vec{q}}/\nu_{\vec{q}})}$ (see e.g.~\cite{cover05}), and the last inequality follows from the nonnegativity thereof.
Consequently, we arrive at
$
(1-\eta(\epsilon))R
\leq
\log{(\sum_qd_q^{\sR})}
$.
Since this relation holds for any $\epsilon>0$ and $R<\mathfrak{C}(\sigma)$, we obtain $\mathfrak{C}(\sigma)
\leq
\log{(\sum_qd_q^{\sR})}
$.

The entropies of the state $\varsigma_n$ are evaluated as follows.
Since $\ca{E}_x(\sigma^{\otm n})$ is a symmetric state, it can be represented as
\alg{
\ca{E}_x(\sigma^{\otm n})
=
\sum_{\vec{q}\in Q^{\times n}}\mu_{\vec{q}|x}\proj{\vec{q}}^{\bsz}\otm\pi_{\vec{q}}^{\bsL}\otm\varsigma_{\vec{q},x}^{\bsR},
\laeq{eq16}
}
where 
$\pi_{\vec{q}}^{\bsL}$ is the maximally mixed state on $\ca{H}_{\vec{q}}^{\bsL}$, 
$\{\mu_{\vec{q}|x}\}_{\vec{q}\in Q^{\times n}}$ is a probability distribution for each $x$
and
$\varsigma_{\vec{q},x}\in\ca{S}(\ca{H}_{\vec{q}}^{\bsR})$.
It follows that
$
H(S^n)_{\ca{E}_x(\sigma^{\otm n})}
=
H(\{\mu_{\vec{q}|x}\}_{\vec{q}})+\!\!\!\sum_{\vec{q}\in Q^{\times n}}\!\!\mu_{\vec{q}|x}[H(\varsigma_{\vec{q},x})
\!+
\log{d_{\vec{q}}^{\bsL}}]
$.
This leads to
$
H(S^n|X)_{\varsigma_n}
\!\!=\!\!
2^{-nR}\sum_{x=1}^{2^{nR}}H(S^n)_{\ca{E}_x(\sigma^{\otm n})}
\!\!\geq\!\!
\sum_{\vec{q}\in Q^{\times n}\!}
\mu_{\vec{q}}\log{d_{\vec{q}}^{\bsL}}
$,
where
$
\mu_{\vec{q}}:=2^{-nR}\sum_{x=1}^{2^{nR}}\mu_{\vec{q}|x}
$.
It also follows from \req{eq15} and \req{eq16} that
$
{\rm Tr}_X[\varsigma_n]=
\sum_{\vec{q}\in Q^{\times n}}\mu_{\vec{q}}\proj{\vec{q}}^{\bsz}\otm\pi_{\vec{q}}^{\bsL}\otm\varsigma_{\vec{q}}'^{\bsR}
$,
where $\varsigma_{\vec{q}}'\in\ca{S}(\ca{H}_{\vec{q}}^{\bsR})$.
This implies that
$
H(S^n)_{\varsigma_n}
=
H(\{\mu_{\vec{q}}\}_{\vec{q}})+\sum_{\vec{q}\in Q^{\times n}}\mu_{\vec{q}}[
\log{d_{\vec{q}}^{\bsL}}
+
H(\varsigma_{\vec{q}}')]
\leq
H(\{\mu_{\vec{q}}\}_{\vec{q}})+\sum_{\vec{q}\in Q^{\times n}}\mu_{\vec{q}}
\log{d_{\vec{q}}^{\bsL}d_{\vec{q}}^{\bsR}}
$.
\QED
\eprf

 \begin{table}[t]
\renewcommand{\arraystretch}{1.5}
  \begin{center}
    \begin{tabular}{|c|c|c|c|} \hline
       \begin{tabular}{c} type of\\ [-2mm] resource \end{tabular}      &   \begin{tabular}{c} encoding\\[-2mm] operations \end{tabular} & $\mathfrak{C}_{\rm free}$   &   \begin{tabular}{c} superdense\\[-2mm] coding \end{tabular} \\ \hline
       \begin{tabular}{c|c} asymmetry & \; \\\cline{1-1}\end{tabular} & & &  \\ 
    Abelian   &    \multirow{3}{*}{\begin{tabular}{c}AN, SP,\;\;\\[-1mm] \;\;SAN, SSP\end{tabular}} & $\log{d_S}$                                                           & $\times$ \\
   irreducible &      & $0$                                                          & $\times$ \\
   otherwise &    & $\in(0,\log{d_S})$                                                        & \checkmark \\ \hline
   athermality & TO     & $0$                                                            & $\times$ \\ \hline
purity & NO     & $0$                                                            & $\times$ \\\hline
 coherence & IO, MIO, SIO                                                          & $\log{d_S}$       & $\times$ \\\hline
magic & SO    & $\log{d_S}$                                                          & $\times$ \\\hline
\multirow{2}{*}{entanglement} & LOCC\:($A\!\leftrightarrow\! B$)     & $\log{d_S}$                                                          & $\times$ \\
	 & LO\:($A$)     & $\log{d_A}$                                                          & \checkmark \\  \hline
    \end{tabular}
  \end{center}
  \caption{Possibility of superdense coding in several resource theories is summarized. NO and SO stand for noisy operations \cite{horodecki2003reversible} and stabilizer operations \cite{veitch2014resource}, respectively, and LOCC for local operations and classical communication.
  The classical information capacity of free states is denoted by $\mathfrak{C}_{\rm free}$, and the total dimension of the system is denoted by $d_S$. Note that $d_S=d_Ad_B$ in the case of the resource theory of entanglement.}
  \label{tb:values}
\end{table}

\section{Superdense Coding in Other Resource Theories}
\lsec{ORT}

We investigate the possibility of superdense coding in other resource theories (see Table \ref{tb:values}).
In any resource theory, states on a system are classified into {\it free states} and {\it resource states}, and operations on the system are into {\it free operations} and {\it non-free operations}.
The minimal assumptions that any resource theory must satisfy are (i) the set of free states is closed under the action of free operations, (ii) the set of free operations are closed under sequential composition, and (iii) any free state can be prepared by a free operation \cite{chitambar2019quantum}.
We may consider a task of encoding classical information into a quantum system under the restriction that the class of operations allowed for encoding is equal to or a subset of the set of free operations.
We could say that superdense coding is possible if the classical information capacity of any free state is a strictly positive constant and that of at least one resource state is strictly larger than that of free states.

Superdense coding is {\it not} possible in any resource theory in which there is only a single free state,
because the capacity of free states is, in that case, equal to zero.
Examples of such resource theories are those of athermality \cite{brandao2013resource,narasimhachar2019quantifying} and purity \cite{horodecki2003reversible}.
The possibility of superdense coding in the resource theories of coherence \cite{streltsov2017colloquium} and magic  \cite{veitch2014resource} depends on the class of operations allowed for encoding.
Superdense coding is {\it not} possible if any free operation is allowed for encoding.
This is because the capacity of any free state is equal to the system dimension and can never be smaller than the capacity of resource states.
In general, superdense coding is {\it not} possible if the number of perfectly distinguishable free states is equal to the system dimension and any free state can be prepared by an operation allowed for encoding.
In that case, however, we may still reconcile the possibility of superdense coding by adopting a strict subset of the set of free operations for encoding.
An example is the resource theory of entanglement:
superdense coding is {\it not} possible if all LOCC operations are allowed for encoding but is possible if the encoding operations are restricted to be local operations on one of the subsystems.

\section{Conclusion} 
\lsec{conc}
We have proposed an analog of superdense coding in the resource theory of asymmetry.
We have proved that superdense coding is possible if and only if the unitary representation of the symmetry is non-Abelian and reducible.
The result provides an information-theoretical classification of symmetries of quantum systems.
Some future directions are to obtain an upper bound on the symmetry-restricted classical information capacity of general states and to extend the result to the one-shot scenario.

\section*{Acknowledgments}
The author thanks Yoshifumi Nakata, Ryuji Takagi and Hayata Yamasaki for useful discussions.
This work is supported by JSPS KAKENHI (Grant No.~18J01329).

\bibliography{bibbib.bib}

\appendix

\section{Capacity under Covariant Encodings}
\lapp{capCE}

An operation $\ca{K}$ on system $S$ is said to be covariant if it satisfies $\ca{K}(U_g(\cdot)U_g^\dagger)=U_g\ca{K}(\cdot)U_g^\dagger$ for any $g\in G$.
In the following, we consider encoding of classical information into a quantum system under the condition that the encoding operations are restricted to be covariant ones.
The following theorem provides conditions for superdense coding to be possible by covariant encodings:

\bthm{densecodingcovariant}
Superdense coding by covariant encodings is possible on system $S$ with symmetry $G$ only if the unitary representation $\mathfrak{U}_G$ of $G$ on $\ca{H}^S$ is non-Abelian and reducible,
and is possible if it holds that
\alg{
\max_{q\in Q}\min\{d_q^{\sL},d_q^{\sR}\}\geq2.
\laeq{kegarenai}
}
\ethm

\bprf
The capacity theorem for symmetric states (Proposition 5 in the main text) is applicable to the case of covariant encodings, because any covariant operation is asymmetry-nongenerating (AN) and the only condition used in the proof of Proposition 5 is that the encoding operations are AN.
The ``only if'' part of \rThm{densecodingcovariant} immediately follows.
To prove the ``if'' part, it suffices to prove that
\alg{
\max_{\rho\in\ca{S}(\ca{H}^S)}\mathfrak{C}(\rho)
\geq
\log{\left(\sum_{q\in Q}d_q^*d_q^{\sR}\right)},
\laeq{MICofSymcov}
}
where $d_q^*:=\min\{d_q^{\sL},d_q^{\sR}\}$.
Note that the R.H.S. of the above inequality is strictly greater than the capacity of symmetric states (Equality \req{MICofSym} in the main text) if the condition \req{kegarenai} is satisfied.
We prove \req{MICofSymcov} by showing that, for any $\rho\in\ca{S}(\ca{H}^S)$, it holds that
\alg{
\!
\mathfrak{C}(\rho)
\geq
H(\{p_q\})\!+\!\sum_{q\in Q}p_q[H(\rho_q^{\sL})\!+\!\log{d_q^{\sR}}]\!-\!H(\rho).\!
\laeq{HHcov}
}
Here, $p_q:={\rm Tr}[\bra{q}^{\sz}\rho\ket{q}^{\sz}]$ and $\rho_q^{\sL}:=p_q^{-1}{\rm Tr}_{\sR}\bra{q}^{\sz}\rho\ket{q}^{\sz}$.
The R.H.S. of the above inequality is equal to $\log{(\sum_{q\in Q}d_q^*d_q^{\sR})}$ for the state
$
\ket{\psi^*}\!=\!\sum_{q\in Q}(d_q^*d_q^{\sR}/d_S')^{1/2}\ket{q}^{\sz}\ket{\psi_q^*}^{\sL\sR},
$
where $d_S'\!:=\!\sum_{q\in Q}d_q^*d_q^{\sR}$ and $\ket{\psi_q^*}\!\in\!\ca{H}_q^{\sL}\!\otm\!\ca{H}_q^{\sR}$ is the maximally entangled state of Schmidt rank $d_q^*$ for each $q$.

The proof of Inequality \req{HHcov} proceeds along the same line as the proof of Proposition 3 in the main text.
Instead of the projector $\Pi_{n,\delta}^0$,
we define
$
\Pi_{n,\delta}'^0:=
\sum_{\vec{q}\in\ca{T}_{n,\delta}}
\proj{\vec{q}}^{\bsz}\otm 
\Pi_{\vec{q},\delta}^{\bsL}\otm I_{\vec{q}}^{\bsR}
$,
where $\Pi_{\vec{q},\delta}$ is the projector onto the conditionally typical subspace of $\ca{H}_{\vec{q}}^{\bsL}$ with respect to the state $\rho_{\vec{q}}^{\bsL}$.
We consider a unitary $W'$ on $S^n$ in the form of
$
W'=\sum_{\vec{q}\in Q^{\times n}}\proj{\vec{q}}^{\bsz}\otm I_{\vec{q}}^{\bsL}\otm v_{\vec{q}}^{\bsR}
$,
which is covariant because $U_{\vec{g}}$ is decomposed into
$
(\Gamma^{\otm n}) U_{\vec{g}}(\Gamma^{\dagger\otm n})=\sum_{\vec{q}\in Q^{\times n}}\proj{\vec{q}}^{\bsz}\otimes u_{\vec{g},\vec{q}}^{\bsL}\otimes I_{\vec{q}}^{\bsR}
$.
We have
$
{\rm Tr}[\Pi_{n,\delta}'^0W'\rho^{\otm n}W'^\dagger]
\geq1-\epsilon
$,
and the averaged state over the unitaries is given by
$
\bar{\rho}_n
:=\mbb{E}_{W'}[W'\rho^{\otm n}W'^\dagger]
=
\sum_{\vec{q}\in Q^{\times n}}p_{q_1}\cdots p_{q_n}\proj{\vec{q}}^{\bsz}\!\otm\! \rho_{\vec{q}}^{\bsL}\!\otm\! \pi_{\vec{q}}^{\bsR}
$.
It follows that
$
\Pi_{n,\delta}'^0\bar{\rho}_n\Pi_{n,\delta}'^0
\leq
2^{-n[\hat{D}'(\rho)-\delta]}
\cdot \Pi_{n,\delta}'^0
$,
where we have denoted the R.H.S. of \req{HHcov} by $\hat{D}'(\rho)$.
By the same argument as in the proof of Proposition 3, there exists a coding protocol with rate $\hat{D}'(\rho)-3\delta$ and error $13\sqrt{\epsilon}$.
Since $\epsilon,\delta>0$ can be arbitrarily small, we complete the proof of Inequality \req{HHcov}.
\QED
\eprf

%

%


\end{document}